\documentclass[lettersize,journal]{IEEEtran}
\usepackage{amsmath,amsfonts}
\usepackage{algorithmic}
\usepackage{algorithm}
\usepackage{array}
\usepackage[caption=false,font=normalsize,labelfont=sf,textfont=sf]{subfig}
\usepackage{textcomp}
\usepackage{stfloats}
\usepackage{url}
\usepackage{placeins}
\usepackage{verbatim}
\usepackage{graphicx}
\usepackage{cite}
\usepackage[dvipsnames]{xcolor}
\usepackage{booktabs}
\usepackage{multirow}
\usepackage{fix-cm}

\usepackage[framemethod=TikZ]{mdframed}
\begin{document}
\bstctlcite{IEEEexample:BSTcontrol}

\title{Do Reviews Matter for Recommendations in the Era of Large Language Models?}


\author{{Chee~Heng~Tan, 
        Huiying~Zheng,
        Jing~Wang,
        Zhuoyi~Lin,
        Shaodi~Feng,
        Huijing~Zhan, 
        Xiaoli~Li,~\IEEEmembership{Fellow,~IEEE,}
        and J.~Senthilnath,~\IEEEmembership{Senior Member,~IEEE.}
}
\thanks{This work has been submitted to the IEEE for possible publication. Copyright may be transferred without notice, after which this version may no longer be accessible.}}

\markboth{Journal of \LaTeX\ Class Files,~Vol.~14, No.~8, August~2021}%
{Shell \MakeLowercase{\textit{et al.}}: A Sample Article Using IEEEtran.cls for IEEE Journals}

\maketitle

 \begin{abstract}
With the advent of large language models (LLMs), the landscape of recommender systems is undergoing a significant transformation. Traditionally, user reviews have served as a critical source of rich, contextual information for enhancing recommendation quality. However, as LLMs demonstrate an unprecedented ability to understand and generate human-like text, this raises the question of whether explicit user reviews remain essential in the era of LLMs. 
In this paper, we provide a systematic investigation of the evolving role of text reviews in recommendation by comparing deep learning methods and LLM approaches.
Particularly, we conduct extensive experiments on eight public datasets with LLMs
and evaluate their performance in zero-shot, few-shot, and fine-tuning scenarios.
We further introduce a benchmarking evaluation framework for review-aware recommender systems, RAREval, to comprehensively assess the contribution of textual reviews to the recommendation performance of review-aware recommender systems. Our framework examines various scenarios, including the removal of some or all textual reviews, random distortion, as well as recommendation performance in data sparsity and cold-start user settings.
Our findings demonstrate that LLMs are capable of functioning as effective review-aware recommendation engines, generally outperforming traditional deep learning approaches, particularly in scenarios characterized by data sparsity and cold-start conditions.
In addition, the removal of some or all textual reviews and random distortion does not necessarily lead to declines in recommendation accuracy.
These findings motivate a rethinking of how user preference from text reviews can be more effectively leveraged. All code and supplementary materials are available at: https://github.com/zhytk/RAREval-data-processing.

\end{abstract}


\section{Introduction}

Recommender systems are widely used in online platforms to predict users’ next actions, such as the next video to watch, the next product to purchase, or items they may prefer based on behavioral histories and item reviews \cite{cremonesi2010performance,lin2021glimg,sun2019research, isinkaye2015recommendation}. 
In recent years, deep learning techniques have shown substantial improvements in modeling user preferences and item properties with textual reviews \cite{zheng2017joint, tay2018multi, lin2022attention, transnets}. 
For example, DeepCoNN \cite{zheng2017joint} uses convolutional neural networks to extract latent features from review texts, delivering more effective recommendation by capturing semantic nuances and latent representations. 
Recent models like RGCL \cite{shuai2022review} and DIRECT \cite{wu2024direct} exploit reviews by respectively combining contrastive learning with Graph Neural Networks (GNNs) and using a concept bottleneck to attribute user preferences to aspect-level item descriptions.

\IEEEpubidadjcol

\begin{figure}[!t]
    \centering
    \vspace{-0.1in}
    \subfloat{%
        \includegraphics[width=0.46\textwidth]{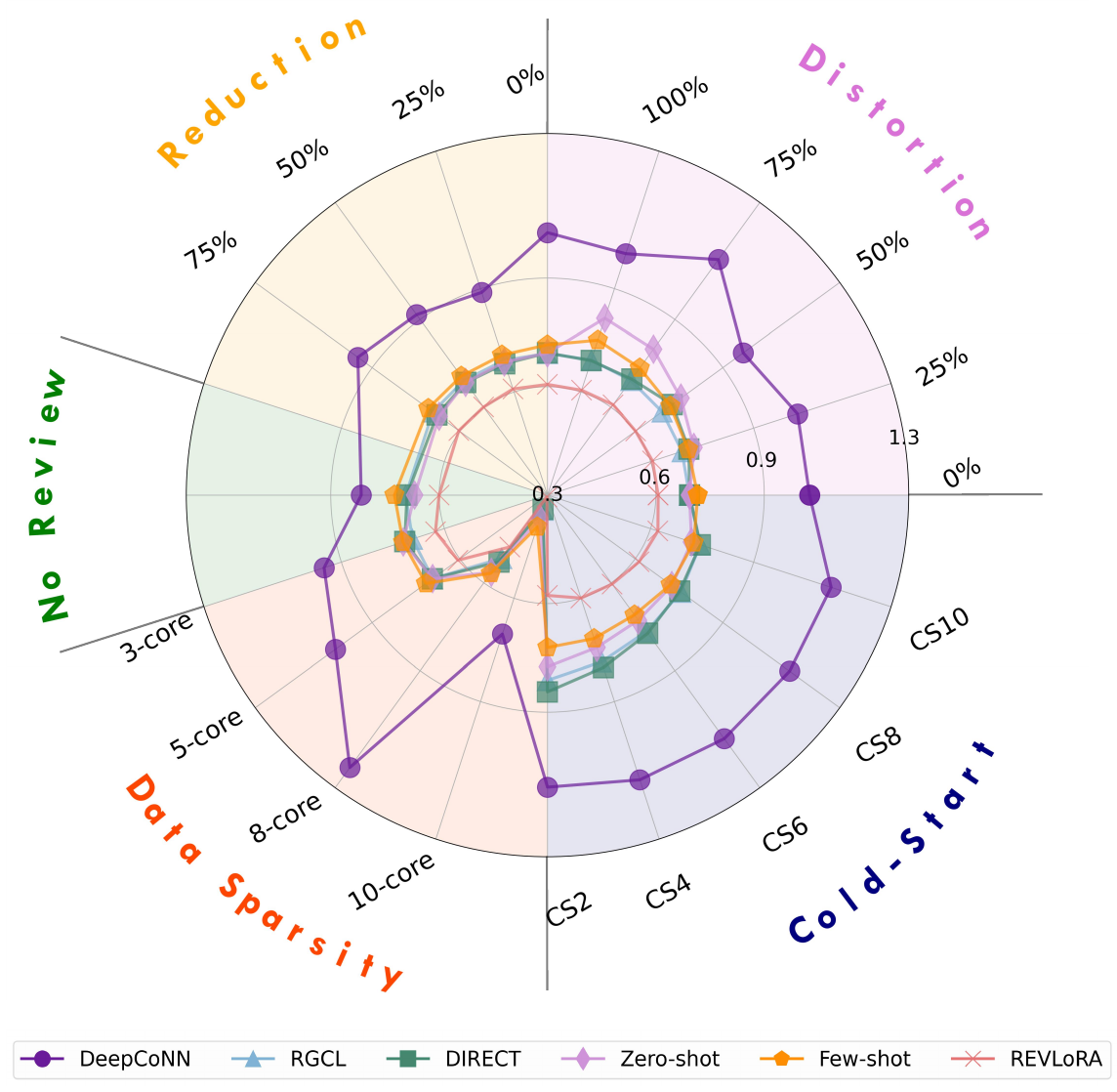} 
    }
    \vspace{-0.05in}
\caption{A comprehensive comparison of recommendation models evaluated using our RAREval framework, which reports the average MAE across Amazon datasets under review removal, distortion, and reduction. RAREval further assesses model performance under data sparsity ($k$-core, where each user and item has at least $k$ reviews) and cold-start scenarios (CS$k$, where users have at most $k$ interactions).}
    \label{fig:all-results }
    \vspace{-0.25in}
\end{figure}

\raggedbottom

On the other hand, Large Language Models (LLMs) have gained significant attention due to their capabilities in understanding complex language patterns, processing large amounts of data, and reasoning through contextual information \cite{touvron2023llama,yenduri2024gpt}. This motivates researchers to propose LLM-based recommendation solutions to predict users' preferences and achieve more effective recommendation \cite{wu2024survey, kang2023llms,geng2022recommendation, zhao2024recommender, li2025personalized}. For instance, Kang et al. systematically evaluate the performance of LLMs for rating prediction tasks, showing great potential in improving data efficiency through fine-tuning \cite{kang2023llms}.
Intuitively, LLMs possess a strong capacity to capture rich, nuanced information from text reviews, which can accurately capture users' preferences and items' properties for modern recommender systems\footnote{In this work, we focus on the rating prediction task rather than sequential recommendation, which imposes strict causality constraints to preserve item order \cite{kang2018selfattentive}. This choice is motivated by three reasons: (1) our objective is to assess the contribution of text reviews to recommendation performance independent of the temporal order of user-item interactions; (2) most of the existing review-based methods are designed for rating prediction, and we adopt this setting for consistency and fair comparison\cite{zheng2017joint, wu2024direct, shuai2022review,mcauley2013hidden,bao2014topicmf}; and (3) applying LLMs to sequential recommendation requires timestamp-aligned item descriptions, complicating both the processing of text reviews and the evaluation of their utility.}. This ability positions LLMs as a promising tool for enhancing the personalization and contextual relevance of review-aware recommendations. 
However, most existing LLM-based recommendation approaches generally overlook textual reviews, relying predominantly on historical user-item interactions. This oversight limits their ability to capture nuanced user preferences and item semantics. Integrating text reviews into LLM-based frameworks offers not only a promising path toward improved personalization and performance in real-world applications but also raises critical questions about how such review-aware LLM solutions should be evaluated. 

In this study, we aim to offer a principled basis for understanding the role of textual reviews in modern, LLM-based recommender systems.
To this end, we undertake a comprehensive empirical investigation into the behavior and effectiveness of deep learning models and LLMs across eight datasets. Specifically, we demonstrate the effectiveness of \emph{\textbf{LLMs in functioning as review-aware recommendation engines}}, systematically evaluating their performance under three learning paradigms: (i) \textbf{zero-shot} learning, wherein the model generates predictions without access to task-specific examples; (ii) \textbf{few-shot} learning, which utilizes a limited number of labeled demonstrations to support LLM inference; and (iii) \textbf{fine-tuning}, wherein LLMs are adapted end-to-end using users’ historical ratings and reviews via our proposed REVLoRA approach. 
Meanwhile, we introduce \emph{\textbf{RAREval, a comprehensive evaluation framework for review-aware recommender systems}}, designed to rigorously assess the contribution of textual reviews to recommendation accuracy.  RAREval examines a range of distinct scenarios, including the removal of some or all text reviews, random distortion of text reviews, and the recommendation performance in various data sparsity levels and cold-start user scenarios. 
In summary, the main contributions of this study are as follows:

\begin{itemize}
    \item To the best of our knowledge,  this is the first systematic study evaluating the contribution of textual reviews to the performance of review-aware recommender systems in the context of LLMs.
    \item We explore LLMs in zero-shot, few-shot, and fine-tuning settings, demonstrating their effectiveness in leveraging text reviews for recommendation, with particular advantages of REVLoRA in mitigating data sparsity and cold-start issues. 
    \item We introduce RAREval, an evaluation framework for analyzing the contribution of text reviews to review-aware recommender systems from multiple perspectives.
    \item We evaluated the performance of deep learning models and LLM-based solutions across eight datasets using the RAREval framework. The results reveal that removing, partially removing, or randomly distorting textual reviews does not consistently diminish recommendation accuracy. These findings prompt a reconsideration of the assumed utility of textual reviews in recommendation tasks and provide a basis for refining practical system design.
\end{itemize}

The rest of the article is organized as follows: We briefly review related works in Section \uppercase\expandafter{\romannumeral2}. Section \uppercase\expandafter{\romannumeral3} presents large language model solutions for review-aware recommendation. in Section \uppercase\expandafter{\romannumeral4}, we introduce the evaluation framework RAREval. In Section \uppercase\expandafter{\romannumeral5}, we empirically evaluate deep learning baselines and the proposed LLM solutions using RAREval.  We conclude the paper and outline directions for future research in \uppercase\expandafter{\romannumeral6}.

\section{Related Work}
In this section, we briefly review the state-of-the-art review-aware recommendation methods. We then discuss how recent advancements leverage LLMs to deliver more contextually rich and personalized recommendations.

\subsection{Review-aware Recommendation}

Early attempts such as HFT \cite{mcauley2013hidden} and TopicMF\cite{bao2014topicmf} are proposed to incorporate reviews by modeling topics alongside latent user and item representations, regularizing the latent representations being learned through Matrix Factorization \cite{koren2009matrix}.
Given the powerful ability of deep learning models to capture fine-grained semantic signals in text reviews, these techniques have been widely adopted in developing recommendation systems to improve recommendation accuracy \cite{fan2022comprehensive}. 
A representative work, DeepCoNN \cite{zheng2017joint}, integrates two parallel convolutional neural networks within a shared final layer, facilitating interaction between latent user and item factors from text reviews.
TransNets \cite{transnets} builds upon DeepCoNN by incorporating a transformation layer that approximates review representations, thereby enabling more personalized recommendations through a more flexible utilization of the data.
Among attention-based models, NARRE \cite{chen2018neural} enhances interpretability by applying single-point attention to highlight key reviews for each user-item pair, while MPCN \cite{tay2018multi} introduces a multi-pointer co-attention mechanism to capture multiple salient interactions, leading to richer and more nuanced recommendations, particularly in semantically complex scenarios. Meanwhile, graph neural network approaches incorporate textual reviews and user-item interactions as essential information sources within graph-based recommendation models. For instance, RGCL \cite{shuai2022review} integrates contrastive learning into graph structures to improve the robustness and discriminative power of learned representations, while REHCL \cite{wang2024enhanced} constructs topic and semantic graphs to capture review relationships from multiple complementary perspectives.
A more recent model, DIRECT \cite{wu2024direct}, is a review-aware recommendation method that employs pre-trained language models to represent textual reviews, enhancing both accuracy and interpretability by modeling user preferences across diverse item aspects.

\subsection{Large Language Models for Recommendation}
The integration of LLMs with recommender systems has garnered significant attention due to their capacity to enhance personalization and process complex, unstructured data, such as user reviews, especially exhibiting the generalization ability in zero-shot and few-shot tasks\cite{wu2024survey}. 
As a preliminary exploration, ChatGPT has demonstrated strong performance in rating prediction tasks under zero-shot or few-shot settings \cite{liu2023chatgptgoodrecommenderpreliminary}. Furthermore, LLMs have shown comparable or even superior performance relative to traditional models while utilizing only a small fraction of the training data, primarily through fine-tuning that substantially improves data efficiency \cite{kang2023llms}.
In conversational recommender systems \cite{sun2018conversational}, LLMCRS \cite{feng2023large} and Chat-REC \cite{gao2023chat} leverage dialogue-based interfaces to deliver high-quality recommendations by addressing tasks including eliciting user preferences, providing personalized suggestions, offering item explanations, and assisting with information retrieval. LLM2ER-EQR \cite{yang2024fine} further advances model transparency through a reinforcement learning-based fine-tuning process combined with an interpretable reward model, which improves the consistency and personalization of generated explanations.

Despite increasing interest in LLM-based recommendation engines, most existing approaches overlook auxiliary information, particularly user reviews when modeling user preferences. 
While several studies have explored related directions, such as P5 \cite{geng2022recommendation}, which incorporates review summarization as a pretraining task for zero-shot recommendation, or efforts that use LLMs to preprocess textual metadata for downstream representation learning \cite{li2023exploring,v2025contrastive} and embedding alignment \cite{ren2024representation}, these works do not position LLMs as the primary recommendation models directly driven by textual reviews. Moreover, they fall short in assessing the actual contribution of reviews to recommendation performance.
Motivated by the above observation, we investigate leveraging LLMs to effectively harness insights from reviews, and integrating unstructured textual data into LLM-based recommender systems.
\begin{figure*}[!ht]
    \centering
    \subfloat{
    \includegraphics[width=0.95\textwidth]{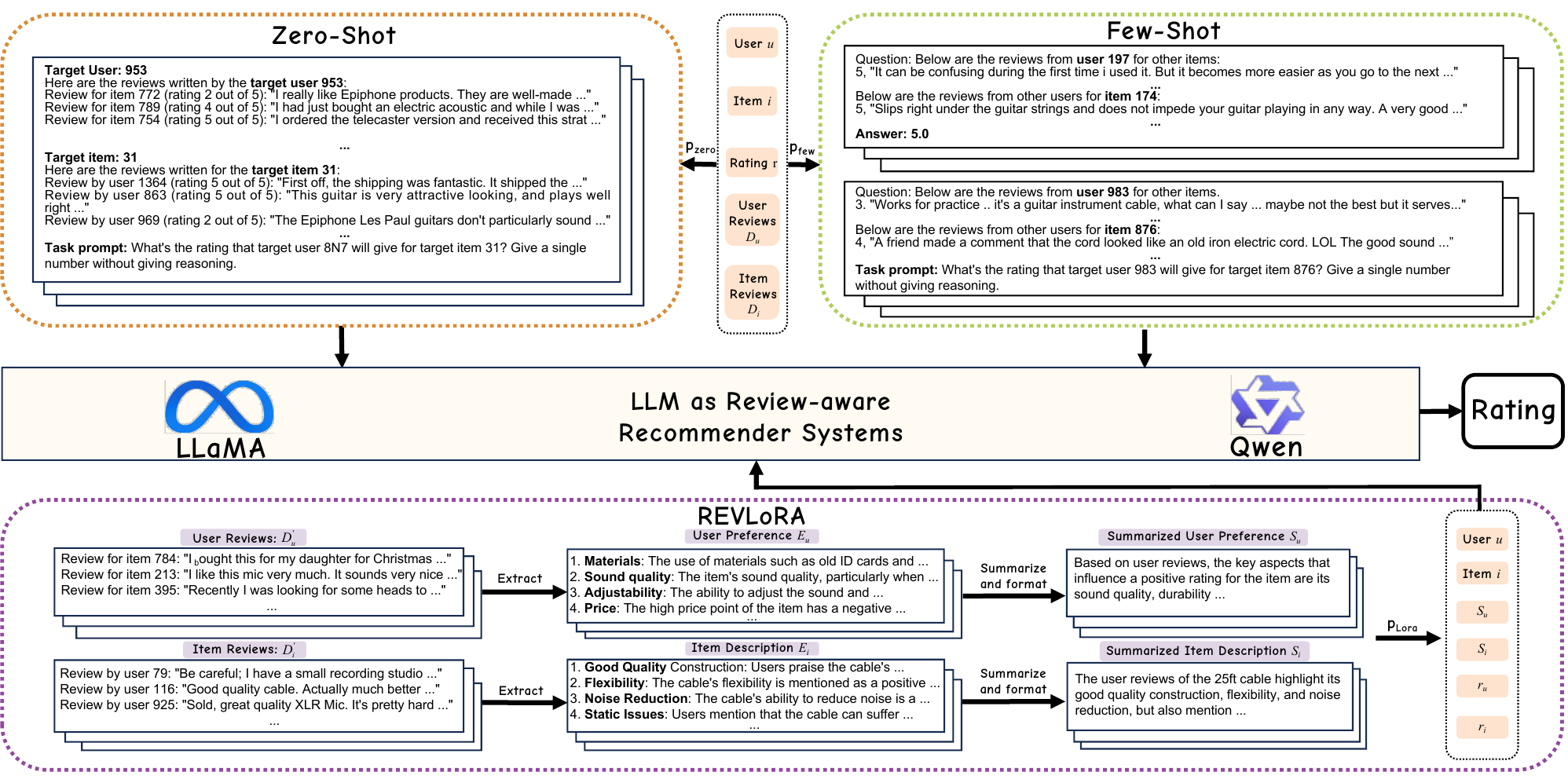}
    }
    \vspace{-3mm}
    \caption{Illustration of prompt formats and workflows for zero-shot, few-shot, and REVLoRA fine-tuning settings.}
\label{fig:prompt} 
    \vspace{-3.5mm}
\end{figure*}
\
\section{Methods}
We systematically explore multiple approaches for leveraging large language models (LLMs) in review-aware rating prediction, with a focus on zero-shot, few-shot, and our proposed fine-tuned method, termed \textbf{Rev}iew-aware \textbf{LoRA} Fine-tuning (REVLoRA). Figure~\ref{fig:prompt} illustrates the prompt structure and workflow corresponding to each of the three settings. In the subsequent subsections, we first define the problem formulation, followed by a detailed description of each approach.

\subsection{Problem Formulation}

Let $\mathcal{U}$ denote the set of users and $\mathcal{I}$ the set of items. For any user $u \in \mathcal{U}$ and integer $k$, let $r^{k}_u$, $R^{k}_u$, and $\text{id}^{k}_u$ denote the rating, review text, and item ID of the $k$-th review written by user $u$, respectively. Similarly, for any item $i \in \mathcal{I}$, let $r^{k}_i$, $R^{k}_i$, and $\text{id}^{k}_i$ represent the rating, text review, and user ID of the $k$-th review associated with item $i$. 
We collate the rating, review text and ID as the user review data $D_u = \{(r^{1}_u, R^{1}_u, id^{1}_u), (r^{2}_u, R^{2}_u, id^{2}_u), ...\}$ and item review data $D_i = \{(r^{1}_i, R^{1}_i, id^{1}_i), (r^{2}_i, R^{2}_i, id^{2}_i), ...\}$. 
Given a user $u$, an item $i$, and their corresponding reviews $D_u$ and $D_i$, the objective is to predict the rating that user $u$ would assign to item $i$.

\subsection{Zero-shot Recommendation}
Zero-shot LLMs have demonstrated impressive reasoning capability with appropriate prompt engineering methods \cite{kojima2023largelanguagemodelszeroshot, chen2023llmarevisualreasoningcoordinators, gruver2023llmarezeroshottimeseriesforecasters}.
For example, zero-shot LLM performs slightly worse than deep learning recommender systems in sequential recommendation tasks \cite{wang2023zeroshot}, while outperforming fine-tuned conversational recommender systems \cite{he2023llmforzeroshotcrs}. 
Nonetheless, it is still uncertain to what extent zero-shot LLMs, when leveraging textual reviews, are capable of reasoning about the subtleties of user preferences and identifying the salient item attributes that appeal to individual users.
Our zero-shot prompt format, as shown in Fig. \ref{fig:prompt} is defined as $p_{zero}(u, i, D_u, D_i)$, where $p_{zero}$ converts the raw data into a prompt by listing the user ID, item ID, the reviews written by the user, followed by the reviews written by the item, and subsequently queries the language model to predict the rating. 

\subsection{Few-shot Recommendation}
Analogously, the few-shot function is defined as $p_{few}(Gen, u, i, D_u, D_i)$, where $Gen$ is a pseudo-random generator that generates the demonstrations. Specifically, we randomly sample historical interactions and their associated reviews using $Gen$, excluding those involving the target user $u$ or the target item $i$. These demonstrations serve to calibrate the model’s rating expectations. The prompt then concludes with a final query that asks the LLM to predict the rating that the target user $u$ would assign to the target item $i$, conditioned on the user’s review data $D_u$ and the item’s review data $D_i$, following a structure analogous to the zero-shot setting. 

\subsection{Review-aware LoRA Fine-tuning (REVLoRA)}
While text reviews are often processed by a concatenation process to construct user profiles and item descriptions \cite{zheng2017joint, wu2024direct, chin2018anr}, it is worth noting that fine-tuning LLMs on such concatenated inputs is generally impractical, even with parameter-efficient fine-tuning techniques such as Low-Rank Adaptation (LoRA) \cite{hu2022lora}. 
Table \ref{tab:concatenation} highlights the inefficiency of fine-tuning the Qwen 0.5B model on the Musical Instruments dataset using the concatenated reviews. This inefficiency becomes more pronounced with larger datasets and model sizes. A key factor contributing to the extended fine-tuning time is the length of the input prompt, which grows significantly when concatenating multiple reviews. To address this challenge, one intuitive solution is to compress the review text while preserving the most relevant information \cite{ren2024representation, v2025contrastive}.

\begin{table}
    \centering
        \caption{Comparison of different reviews processing solutions on the Musical Instruments (5-core) dataset.  Qwen 0.5B is fine-tuned with LoRA using a single NVIDIA RTX A5000 GPU. REVLoRA improves efficiency and effectiveness with both extraction and summarization.}
    \begin{tabular}{m{1.7cm} | m{0.8cm} | m{1.58cm} | m{1.35cm} | m{1.45cm}}
    
        \toprule
        \midrule
        
         Method&  MAE   & Preprocessing Time (min) & Fine-tuning Time (min) & Runtime Penalty\\

        \midrule
         Concatenation&  0.5422   & N/A & 3,246 & 13.89$\times$ \\ 
                 \midrule
         Extraction &  0.5334   & 151 & 624 & 2.56$\times$ \\
                 \midrule
         REVLoRA&  0.5338   & 171 & 47 & N/A \\         
        \midrule
        \bottomrule           
    \end{tabular}
    \label{tab:concatenation}
    \vspace{-0.15in}
\end{table}

To this end, we introduce a novel solution named \textbf{Rev}iew-aware \textbf{LoRA} Fine-tuning (REVLoRA) to fine-tune the LLMs more efficiently. The three stages are in order: extraction, summarization\footnote{The extraction and summarization steps could be combined via chain-of-thought \cite{wei2023chainofthoughtpromptingelicitsreasoning} or by using larger LLMs. We leave this to further studies.}, and fine-tuning. As evidenced in Table \ref{tab:concatenation}, the time savings in fine-tuning shorter prompts far outweigh the time spent extracting and summarizing the review texts, making REVLoRA more computationally efficient.
Let $D'_{u}=\{(R_{u}^{k}, id_{u}^{k})| (r_{u}^{k}, R_{u}^{k}, id_{u}^{k}) \in D_u\}$ be the set of user review text data with the corresponding item IDs and $D'_{i}=\{(R_{i}^{k}, id_{i}^{k})| (r_{i}^{k}, R_{i}^{k}, id_{i}^{k}) \in D_i\}$ be the set of item review data with the corresponding user IDs. 
As depicted in Figure \ref{fig:prompt}, we used LLM to extract features of the user $E_u = e_u(D'_{u})$ and the item $E_i = e_i(D'_{i})$ that influence the rating that the user will give to the item, where $e_u$ and $e_i$ are user and item feature extraction functions respectively. 
However, fine-tuning using the extracted user and item features $E_u$ and $E_i$ remains computationally impractical, as shown in Table \ref{tab:concatenation}. 
Consequently, we further summarize $E_u$ and $E_i$ to sentence-long user preferences $S_u$ and item descriptions $S_i$ respectively, for each $u \in U$ and $i \in I$, i.e. $S_u = s_u(E_u)$ and $S_i = s_i(E_i)$ for summarization functions $s_u$ and $s_i$. 
Following the extraction and summarization processes, we construct the fine-tuning prompt and fine-tune the LLM using LoRA. Where $r_u=\{r_u^1, r_u^2,...\}$ and $r_i=\{r_i^1, r_i^2,...\}$ are the sets of user and item ratings respectively, the LoRA prompt $p_{LoRA}(u, i, S_u, S_i, r_u, r_i)$ lists down the user preferences and item description after the summarization stage, followed by the set of user and item ratings. We then adapt the LoRA architecture to suit our task of rating prediction. Following \cite{kang2023llms}, we replace the text generation inference head of the LLM with a single linear layer with the input dimension equal to the token embedding dimension of the LLM, and the output being the rating prediction. Note that all linear layers and the token embedding layers are fine-tuned because fine-tuning them empirically improves the overall performance.

\begin{figure*}[t]
    \centering
    \subfloat{%
        \includegraphics[width=1\textwidth, ]{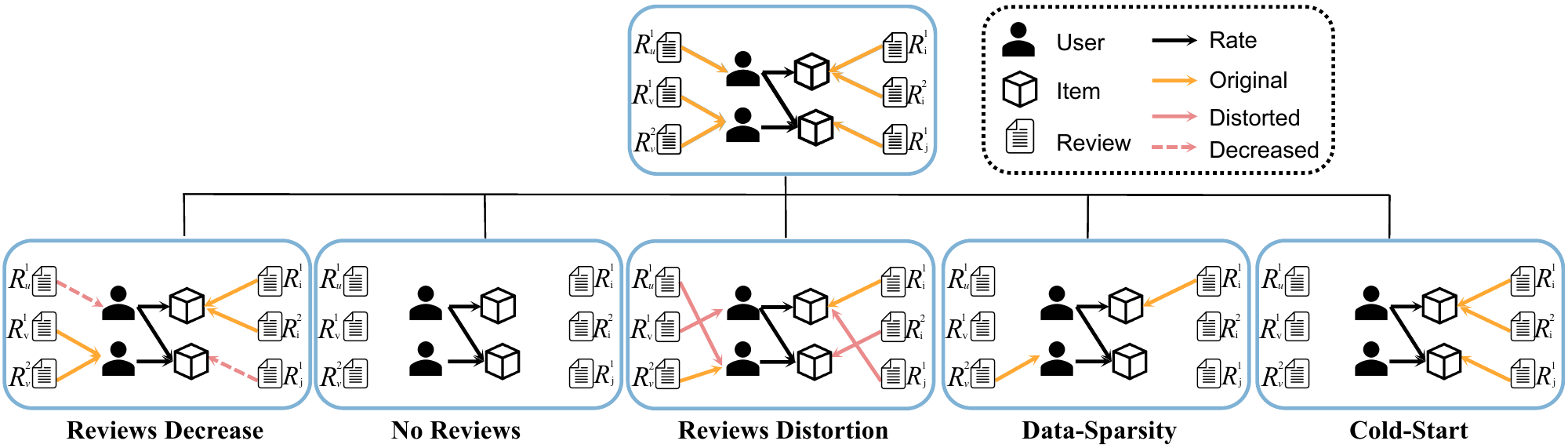}  
    }
    \caption{The RAREval framework evaluates review-aware recommender systems across five distinct settings, each derived from the original dataset.}
    \label{fig:mae_comparison}
        \vspace{-0.15in}
\end{figure*}

\section{Evaluation Framework for Review-aware Recommender Systems}

While several evaluation frameworks have been proposed for recommender systems, including KG4RecEval for knowledge graph-based methods \cite{zhang2025kg4receval} and CRS-Que for conversational recommender systems emphasizing user experience \cite{jin2024usercentricevaluationframework}, evaluation of text reviews has received less attention. While a previous study investigated review usefulness for deep learning models \cite{sachdeva2020useful}, there remains a need for a more comprehensive evaluation framework, particularly in the current era of LLMs.
In this work, we introduce five research questions (RQ1 to RQ5) to systematically evaluate the impact of review data on recommendation performance, each aligned with a key experimental scenario. 

Overall, RQ1, RQ2, and RQ3 investigate whether the performance of review-aware recommender systems is affected when review data is entirely removed, partially reduced, or randomly distorted, respectively. RQ4 extends this analysis to scenarios with varying degrees of data sparsity, examining how the availability of user-item interactions and reviews influences the effectiveness of recommendation performance. Lastly, RQ5 focuses on cold-start users, for whom textual reviews are expected to serve as a primary signal in the absence of sufficient historical interaction data.
These questions guide the evaluation of deep learning and LLM-based methods, enabling a comprehensive analysis of their strengths and limitations in utilizing text reviews. In the following, we detail the experimental settings employed to address each question.

\subsection{\textbf{What if there are no reviews? (RQ1)}}
This experiment evaluates the impact of removing all review data on recommendation performance. We exclude reviews from all phases while retaining the rating information. This review-free setting serves as a baseline to assess model performance without textual input and to quantify the contribution of reviews to recommendation accuracy.

\subsection{\textbf{What if a subset of reviews are removed? (RQ2)}}
While RQ1 investigates the extreme case where no reviews are available, this experiment explores how recommendation performance varies as the proportion of retained reviews decreases. We systematically remove 0\%, 25\%, 50\%, 75\%, and 100\% of reviews in the training set, chosen at random, while keeping the validation and test sets intact. This allows us to assess model robustness to review sparsity by analyzing performance degradation under reduced review availability.

\subsection{\textbf{What if the reviews are distorted? (RQ3)}}
For RQ3, we aim to examine review quality by introducing textual noise. We randomly shuffle reviews among user-item pairs, preserving the original ratings and interaction data. Distortion levels of 0\%, 25\%, 50\%, 75\%, and 100\% are evaluated to assess model robustness to misaligned textual content. This setup helps determine the extent to which recommendation models rely on the alignment between reviews and user-item pairs for effective performance.

\subsection{\textbf{What if the dataset is more or less sparse? (RQ4)}}
Beyond review content, overall data sparsity remains a key challenge in recommender systems. This experiment investigates the impact of varying user-item interaction sparsity on model performance. We apply $k$-core filtering to retain only users and items with at least $k$ interactions, using $k \in \{0, 3, 5, 8, 10\}$. Note that the maximum $k$ varies by dataset due to size constraints.
For example, Musical Instruments supports up to 5-core, and Amazon Instant Video up to 8-core. This setup enables us to assess whether LLM-based methods demonstrate greater robustness under increasing sparsity compared to deep learning models.

\subsection{\textbf{What if the user is in a cold-start situation? (RQ5)}}
Cold start users, who have limited interaction history, present a significant challenge for recommender systems \cite{yuan2023usercoldstart, sun2019research, burke2011recommender}. To evaluate model performance under cold start conditions, we segment users based on their interaction frequency in the training set and evaluate the recommendation performance for each group.
We focus on users with one to ten interactions, as this range reflects limited behavioral data and serves as a practical approximation of cold-start conditions. This experiment evaluates the robustness of deep learning recommendation methods and LLM-based approaches in handling cold-start users.

\begin{table*}[!t]
\caption{Dataset Statistics across different k-core thresholds.}
\label{tab:dataset}
\centering
\scriptsize
\renewcommand\arraystretch{1.2} 
\resizebox{\textwidth}{!}{        
\begin{tabular}{l|ccc|ccc|ccc|ccc|ccc}
\toprule
\midrule
Dataset
& \multicolumn{3}{c|}{0-core} 
& \multicolumn{3}{c|}{3-core}
& \multicolumn{3}{c|}{5-core} 
& \multicolumn{3}{c|}{8-core}
& \multicolumn{3}{c}{10-core} \\
& \# Users & \# Items & \# Reviews 
& \# Users & \# Items & \# Reviews 
& \# Users & \# Items & \# Reviews 
& \# Users & \# Items & \# Reviews 
& \# Users & \# Items & \# Reviews \\
\midrule
Musical Instruments 
& 338,967 & 83,025 & 499,730 
& 17,472 & 9,225 & 82,714
& 1,429 & 900 & 10,261 
& N.A. & N.A. & N.A. 
& N.A. & N.A. & N.A. \\
Amazon Instant Video 
& 426,910 & 23,962 & 583,914 
& 26,676 & 5,525 & 117,555 
& 5,130 & 1,685 & 37,126 
& 418 & 277 & 4,675 
& N.A. & N.A. & N.A. \\
Office Products 
& 909,314 & 130,006 & 1,242,911 
& 40,772 & 15,051 & 197,803 
& 4,905 & 2,420 & 53,237 
& 1,838 & 846 & 30,383 
& 1,380 & 700 & 25,367 \\
Digital Music 
& 478,201 & 266,393 & 835,953 
& 24,645 & 13,526 & 151,338
& 5,541 & 3,568 & 64,706
& 1,841 & 1,544 & 33,781 
& 1,049 & 990 & 22,772 \\
Video Games 
& 826,615 & 50,207 & 1,324,449 
& 78,913 & 21,315 & 442,609
& 24,303 & 10,672 & 231,780
& 6,505 & 4,119 & 98,028
& 2,815 & 2,140 & 52,137 \\
Health and Personal Care 
& 1,851,132 & 252,331 & 2,981,863 
& 168,945 & 55,034 & 870,644 
& 38,609 & 18,534 & 346,307 
& 4,072 & 2,427 & 78,489 
& 2,184 & 1,260 & 55,070 \\
CDs and Vinyl 
& 1,578,597 & 486,360 & 3,748,825 
& 207,476 & 145,490 & 1,779,060 
& 75,258 & 64,443 & 1,097,577 
& 26,553 & 26,051 & 615,722 
& 15,592 & 16,184 & 445,408 \\
Movies and TV 
& 2,088,620 & 200,941 & 4,606,671 
& 304,265 & 81,948 & 2,393,655 
& 123,960 & 50,052 & 1,697,471 
& 51,626 & 29,187 & 1,171,270 
& 33,326 & 21,901 & 958,952 \\
\midrule
\bottomrule
\end{tabular}}
\vspace{-0.1in}
\end{table*}

\begin{table*}[!ht]
\centering
\fontsize{5.7pt}{0.25cm}\selectfont
\caption{Experimental results on Amazon 5-core datasets. The best performing methods are bold, and the second best results are underlined. We conduct experiments using Llama 3.2 and Qwen 2.5. Note that Qwen 0.5B refers to Llama-Qwen 0.5B, which utilizes the Llama 1B model to generate summarizations for subsequent fine-tuning based on Qwen 0.5B. In addition, Llama 1B (MSE) means the model is fine-tuned with MSE loss while Llama 1B and Qwen 0.5B are fine-tuned with MAE loss.} \label{main_performance}
\begin{tabular}{l|c|ccc|cccc|ccccc|ccc}
\toprule
\multicolumn{1}{l|}{\multirow{8}{*}{Dataset}}   & \multicolumn{1}{c|}{\multirow{8}{*}{Metric}}  & \multicolumn{3}{c|}{Deep Learning Methods} & \multicolumn{4}{c|}{Zero-Shot LLMs} & \multicolumn{5}{c|}{Few-Shot LLMs} & \multicolumn{3}{c}{Fine-Tuned LLMs (REVLoRA)} \\
 & & \rotatebox{90}{DeepCoNN} & \rotatebox{90}{RGCL} & \rotatebox{90}{DIRECT} & \rotatebox{90}{Llama 1B} & \rotatebox{90}{Llama 3B} & \rotatebox{90}{Qwen 0.5B} & \rotatebox{90}{Qwen 3B} & \rotatebox{90}{LURP} & \rotatebox{90}{Llama 1B} & \rotatebox{90}{Llama 3B} & \rotatebox{90}{Qwen 0.5B} & \rotatebox{90}{Qwen 3B} & \rotatebox{90}{Llama 1B} & \rotatebox{90}{Qwen 0.5B } &\rotatebox{90}{Llama 1B (MSE) }\\
\midrule
Musical Instruments & MAE & 1.1885 & 0.6443 & 0.6263 & 0.5858 & 0.5575 & 0.6330 & 0.5676 & 0.5411 & 0.6735 & \underline{0.5312} & 0.5560 & 0.5609 & \textbf{0.5271} & 0.5473 & 0.6596\\
Amazon Instant Video & MAE & 0.9331 & 0.6929 & 0.6938 & 0.7425 & 0.7158 & 0.7420 & 0.7037 & 0.7425 & 0.8558 & 0.7439 & 0.7113 & 0.7257 & \textbf{0.6204} & \underline{0.6270} & 0.7012\\
Office Products & MAE & 0.9240 & 0.6297 & 0.6315 & 0.6412 & 0.6305 & 0.6431 & 0.6467 & 0.6455 & 0.7517 & 0.6486 & 0.6401 & 0.6544 & \textbf{0.5557} & \underline{0.5572} & 0.6454 \\
Digital Music & MAE & 0.9547 & 0.6432 & 0.6520 & 0.7266 & 0.6572 & 0.6977 & 0.6899 & 0.7342 & 0.8277 & 0.7174 & 0.7314 & 0.6855 & \textbf{0.5628} & \underline{0.5848} & 0.6707 \\
Video Games & MAE & 1.0263 & 0.7905 & 0.8020 & 0.8795 & 0.8417 & 0.8201 & 0.8693 & 0.8766 & 1.1102 & 0.8709 & 0.8556 & 0.8525 & \textbf{0.7153} & \underline{0.7171} & 0.7971 \\
Health and Personal Care & MAE & 0.8405 & 0.7438 & 0.7592 & 0.7414 & 0.7243 & 0.7395 & 0.7520 & 0.7401 & 0.9230 & 0.7396 & 0.7175 & 0.7743 & \textbf{0.6422} & \underline{0.6556} & 0.7716\\
CDs and Vinyl & MAE & 0.8831 & 0.6534 & 0.6941 & 0.7068 & 0.6728 & 0.6842 & 0.6992 & 0.7075 & 0.7803 & 0.6953 & 0.7139 & 0.6915 & \textbf{0.5749} & \underline{0.5903} & 0.6731\\
Movies and TV & MAE & 0.9303 & 0.7335 & 0.7640 & 0.8466 & 0.7931 & 0.7896 & 0.8130 & 0.8489 & 0.9507 & 0.8342 & 0.8332 & 0.8019 & \textbf{0.6613} & \underline{0.6672} & 0.7525\\
\midrule
Musical Instruments & MSE & 1.8998 & 0.7704 & \textbf{0.7252} & 1.1389 & 1.0877 & 0.9374 & 0.9343 & 1.0556 & 1.4903 & 1.0556 & 1.0514 & 1.0792 & 0.9233 & 1.0280 & \underline{0.7377}\\
Amazon Instant Video & MSE & 1.3132 & 0.9340 & \textbf{0.8810} & 1.5707 & 1.4559 & 1.2003 & 1.2922 & 1.6119 & 1.8853 & 1.6760 & 1.3913 & 1.4912 & 1.0216 & 1.0171 & \underline{0.9105}\\
Office Products & MSE & 1.2262 & \underline{0.7381} & \textbf{0.7360} & 1.1684 & 1.1007 & 1.0131 & 1.0400 & 1.2227 & 1.5181 & 1.2315 & 1.1357 & 1.1588 & 0.8949 & 0.8475 & 0.7464\\
Digital Music & MSE & 1.3715 & 0.7975 & \underline{0.7854} & 1.5296 & 1.2577 & 1.2004 & 1.2286 & 1.5572 & 1.8138 & 1.5620 & 1.4103 & 1.3574 & 0.9395 & 0.8610 & \textbf{0.7850}\\
Video Games & MSE & 1.7449 & \underline{1.0924} & \textbf{1.0880} & 1.9309 & 1.7928 & 1.6119 & 1.7219 & 2.0088 & 2.7193 & 2.0125 & 1.8501 & 1.8064 & 1.3122 & 1.2450 & 1.1026\\
Health and Personal Care & MSE & 1.1489 & \textbf{1.0344} & \underline{1.0399} & 1.7042 & 1.5993 & 1.4158 & 1.5073 & 1.7039 & 2.3075 & 1.7451 & 1.5856 & 1.7276 & 1.2770 & 1.2981 & 1.0677\\
CDs and Vinyl & MSE & 1.2397 & \textbf{0.8365} & 0.8896 & 1.5887 & 1.4323 & 1.3339 & 1.3767 & 1.5630 & 1.7915 & 1.5826 & 1.5067 & 1.4880 & 1.0634 & 1.0110 & \underline{0.8688}\\
Movies and TV & MSE & 1.4043 & \textbf{0.9926} & 1.0241 & 1.9634 & 1.7480 & 1.5327 & 1.6446 & 1.9548 & 2.2588 & 1.9826 & 1.7895 & 1.7529 & 1.1636 & 1.1431 & \underline{1.0169}\\
\bottomrule
\end{tabular}
\vspace{-0.15in}
\end{table*}

\section{Experimental Results}
\subsection{Experimental Settings}

\smallskip\noindent\textbf{Datasets:}
To investigate the effect of dataset size on the performance of the model, we chose the Amazon 2014 (5-core) datasets\footnote{\url{https://cseweb.ucsd.edu/~jmcauley/datasets/amazon/links.html}}\cite{mcauley2015image,he2016ups}. Table \ref{tab:dataset} shows the statistics of the Amazon 2014 datasets.
Following existing work\cite{transnets, shuai2022review, tay2018multi}, we removed all empty reviews in the dataset, and randomly split the review data into training, validation, and test sets in an 80:10:10 ratio. 
During preprocessing, we use Llama 1B \cite{grattafiori2024llama3herdmodels} to generate user profile $S_u$ and item profile $S_i$ based solely on reviews from the training set, for every $u \in U$ and $i \in I$. Specifically, up to 15 reviews are summarized into a single profile for each user and each item. Up to 32 user ratings and 32 item ratings are used for each user-item interaction for rating prediction.
Following  \cite{transnets, tay2018multi}, we ensure that the validation and test phases use only profiles derived from the training set, thereby preventing review-wise data leakage and preserving evaluation integrity.

\smallskip\noindent\textbf{Baselines:} We compare with the following state-of-the-art counterparts
(1) \textbf{DeepCoNN\cite{zheng2017joint}}: A classic deep learning model that utilizes user and item review texts to improve recommendation systems by capturing user preferences and item attributes through deep learning techniques. 
(2) \textbf{RGCL\cite{shuai2022review}}: A graph neural network that combines review data and graph signals, focusing on both explicit interactions and the integration of latent semantic information. 
(3) \textbf{DIRECT \cite{wu2024direct}}: A recent review-aware recommendation system leverages pre-trained language models to encode textual reviews, enabling more accurate and interpretable recommendations by capturing user interests across multiple aspects of items. 
(4) \textbf{LURP \cite{kang2023llms}}: It explores how to adapt LLMs for rating prediction. We reproduce their few-shot approach by omitting text reviews, with Llama 1B as the LLM.

\smallskip\noindent\textbf{Training Details:}
We reproduced the deep learning methods as follows: For DeepCoNN, up to 10 reviews from each user and item were sampled and truncated to a maximum length of 40 tokens. The model was trained for 20 epochs using a batch size of 128, a learning rate of 0.002, and a decay factor of 1. We implemented RGCL using Adam \cite{kingma2017adam} optimizer with a learning rate of 0.01, a review embedding size of 32, and a dropout rate of 0.7. These settings were empirically found to yield the best performance.
DIRECT is trained with AdamW \cite{loshchilov2019decoupled} with a learning rate of $10^{-3}$, a batch size of 32 and a dropout of 0.3. 
We evaluate the LLM on zero-shot, few-shot, and REVLoRA settings. The LLMs used are Llama 3.2 \cite{grattafiori2024llama3herdmodels} and Qwen 2.5 \cite{qwen2025qwen25technicalreport}. 
For zero-shot experiments, we use 10 user and 10 item reviews. For few-shot experiments, we use 6 user and 6 item reviews, drawn from 3 examples (3-shot) from the training set, each with 2 user and 2 item reviews. The maximum token count for text reviews is 7680 in both settings, ensuring a fair comparison based on context window usage.
For LURP\cite{kang2023llms}, we reproduce the few-shot \textit{rating-only} setup, where all review texts are omitted and only rating information is used. The backbone language model is Llama 1B. All other training configurations (e.g., prompt template, token length limits, and sampling strategy) remain consistent with our standard few-shot setting to ensure fair comparison.
For REVLoRA, we use the AdamW \cite{loshchilov2019decoupled} optimizer with a learning rate is $10^{-5}$ for the non-embedding layers and $10^{-6}$ for the embedding layers. Meanwhile, we use a batch size of 64, with a LoRA rank and LoRA alpha both set to 64. We observe that larger datasets tend to reach optimal performance with a smaller epoch count during training. For datasets with up to 80,000 reviews, evaluation on the validation set takes place every epoch, with an early stopping patience of 3 epochs. For datasets with more than 80,000 reviews, evaluation on the validation set takes place every 1,000 steps, with an early stopping patience of 3,000 steps. 

\smallskip\noindent\textbf{Evaluation Metrics:} 
In this study, we use Mean Absolute Error (MAE) to measure the average absolute difference, and Mean Squared Error (MSE) to quantify the average squared difference between predicted and actual ratings.

\subsection{Performance Comparison}

The experimental results presented in Table \ref{main_performance} reveal nuanced performance differences between deep learning methods and LLM solutions in predicting user ratings on Amazon 5-core datasets. Fine-tuned LLMs consistently achieve superior MAE across various product categories. REVLoRA with MAE loss optimization has the best overall MAE performance. However, in terms of MSE, DIRECT and RGCL often outperform LLM-based approaches, including the REVLoRA method fine-tuned with an MSE loss function (i.e., Llama 1B (MSE)). 

REVLoRA underperforms RGCL on large datasets but outperforms it on smaller ones, likely due to RGCL’s graph-based architecture \cite{shuai2022review}, which captures higher-order interactions and tends to overfit on smaller datasets. While REVLoRA underperforms DIRECT on smaller datasets where the use of shopping history data \cite{wu2024direct} boosts DIRECT’s performance, REVLoRA outperforms on larger datasets, as DIRECT prioritizes interpretability over predictive accuracy.

Analyzing prompting strategies, we observe that few-shot LLMs do not consistently outperform zero-shot counterparts, particularly for the Llama series (e.g., Llama 1B and 3B). In contrast, zero-shot and few-shot Qwen models exhibit comparable performance in both settings. This result suggests that the effect of fewer target user and target item reviews in the few-shot setting could outweigh the effect of the presence of the provided examples within a given context window.
Furthermore, scaling from Llama 1B to 3B significantly reduces MAE, underscoring the benefits of increased model size. But scaling from Qwen 0.5B to Qwen 3B does not necessarily reduce MAE. Among larger models, Llama 3B generally outperforms Qwen 3B in zero-shot settings, while both perform comparably in the few-shot configuration. 

The remaining subsections provide a more detailed analysis of the impact of reviews. For each of the three LLM settings, we select the best-performing models: the Llama 3B model for the zero-shot and few-shot settings, and the Llama 1B model for the fine-tuning setting (REVLoRA). 

\begin{figure*}[!ht]
    \centering
    \subfloat{%
    \includegraphics[width=1\textwidth]{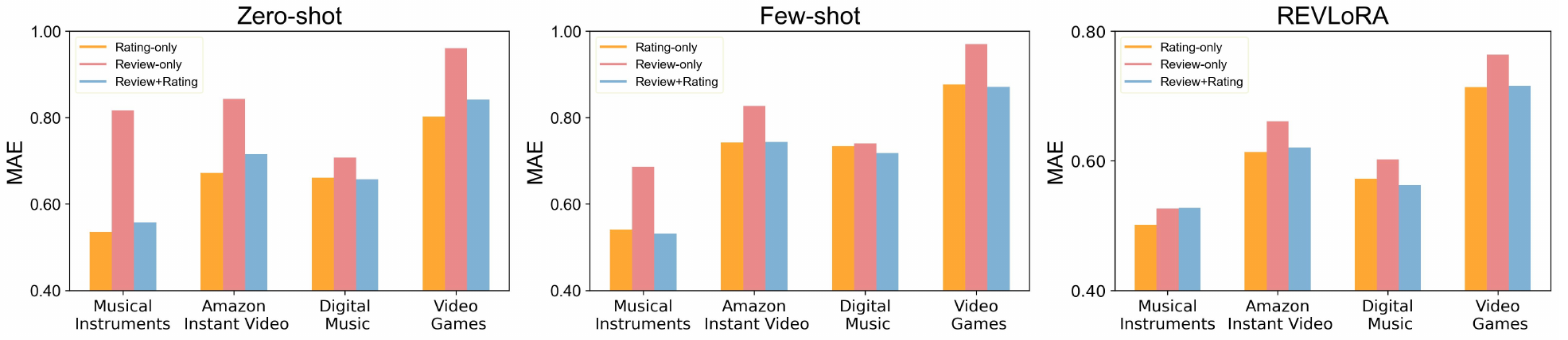} 
    }
        \vspace{-0.1in}
    \caption{
MAE was evaluated on four Amazon datasets using Llama 3B model for Zero-shot and Few-shot settings, and the Llama 1B model with LoRA for the finetuning setting (i.e., REVLoRA). 
}
    \label{fig:ablation-study}
    \vspace{-0.15in}
\end{figure*}

\subsection{The Effect of Reviews}

This subsection evaluates the utility of textual reviews in LLM-based recommendation by isolating ratings and reviews to assess their individual contributions to model performance. We consider three settings: using only rating information (Rating only), only textual reviews (Review only), and both information combined (Review and Rating). MAE results across four benchmark datasets are shown in Fig.~\ref{fig:ablation-study}. 

The LLM in the review-only setting performs consistently worse in the zero-shot, few-shot and REVLoRA settings compared to the rating-only setting, though the review-only performance is decent. Counterintuitively, combining ratings and reviews does not necessarily improve model performance relative to the rating-only setting. 
Although review texts may inherently convey less predictive information than explicit ratings, a more likely explanation is that, within our current experimental setup, LLMs are not fully able to extract or utilize the information embedded in the text reviews.

\begin{mdframed}[linecolor=black, backgroundcolor=gray!10, roundcorner=5pt, innertopmargin=5pt, innerbottommargin=5pt] \textbf{Effect of Review Texts.} 
LLMs can effectively operate as review-aware recommendation engines. LLM demonstrates superior performance in rating-only scenarios compared to review-only settings under our setting, suggesting LLM does not appear to fully leverage the informational richness of review texts.
\end{mdframed}

\begin{figure}[!t]
    \centering
    \subfloat{%
        \includegraphics[width=0.5\textwidth]{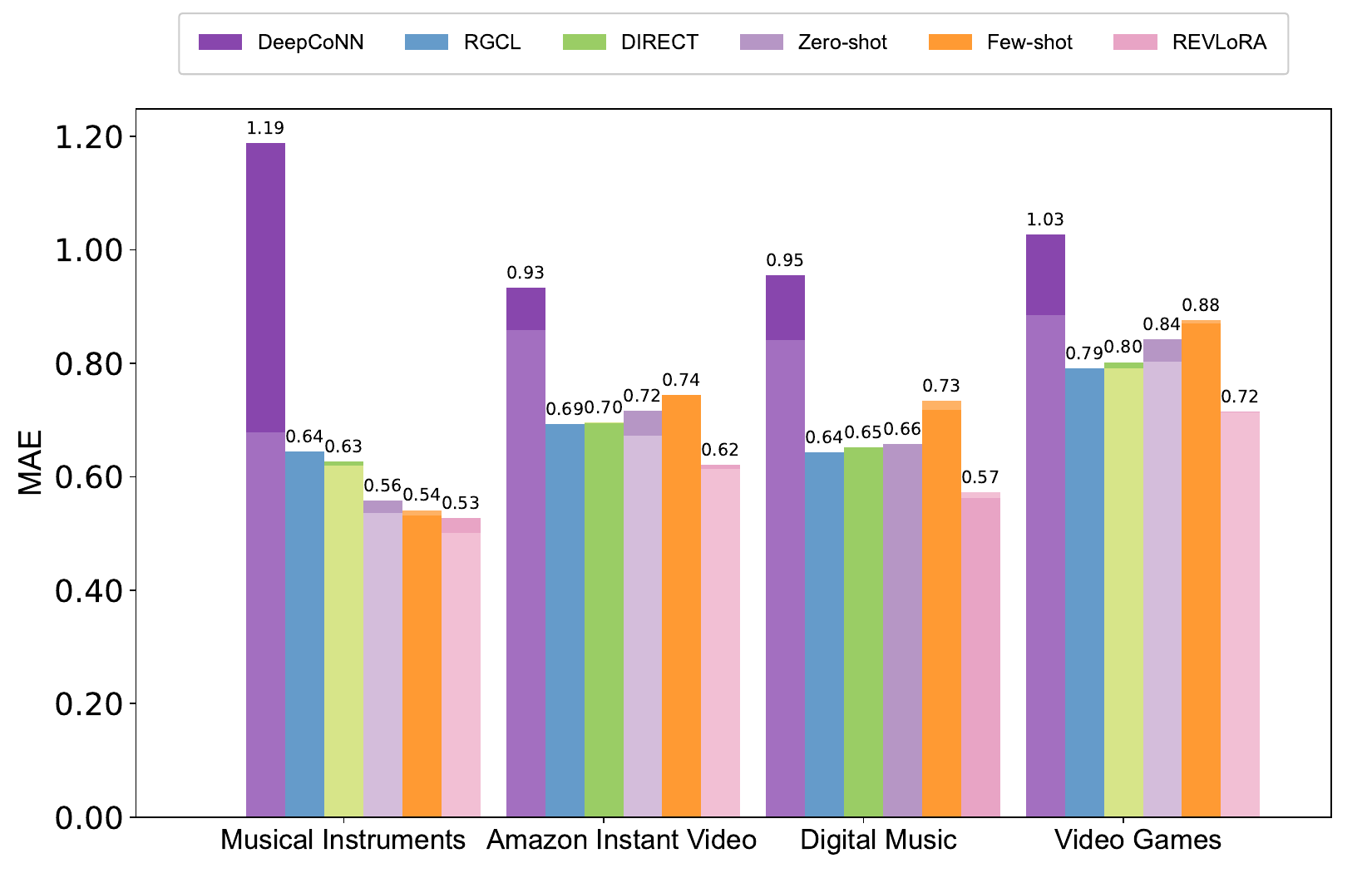} 
    }
    \vspace{-0.1in}
    \caption{MAE results on four datasets. Dark bars denote the default settings (i.e., with review), while light bars indicate performance without reviews.} 
    \label{fig:no review}
        \vspace{-0.15in}
\end{figure}

\subsection{No Reviews Experiment (RQ1)}
To answer RQ1, we conduct an ablation study for both deep learning baselines and LLM solutions by removing all review text during both training and inference stages. Fig. \ref{fig:no review} presents the MAE results across four benchmark datasets. 

DeepCoNN shows counterintuitive performance gains without review content. For example, by removing the review text, MAE improves by 0.51 on the Musical Instruments dataset. This suggests high sensitivity to textual noise, possibly due to limited embedding capacity or aggressive sequence truncation.
In contrast, RGCL shows negligible change, reflecting its reliance on structured ratings and independence from textual features. 
DIRECT shows similar performance with and without reviews, showing that the use of word mention aspects and word sentiments in the model architecture \cite{wu2024direct} did not significantly aid performance.
Zero-shot LLM generally performs worse with reviews, showing that the review text is likely a distractor. This is likely due to the presence of contradictory reviews in the dataset, such as a review that comes with some complaints but with corresponding rating of 5, such as Table 8, Case 3 of \cite{wu2024direct}. The handling of such reviews is complex and likely requires some fine-tuning to account for them. Few-shot LLM generally improves slightly with reviews, likely due to the calibration examples that helped the LLM set rating expectations. REVLoRA exhibits robust performance across datasets, showing generally minimal degradation even in the absence of textual input.

\begin{mdframed}[linecolor=black, backgroundcolor=gray!10, roundcorner=5pt, innertopmargin=5pt, innerbottommargin=5pt] \textbf{Answer to RQ1.} 
Counterintuitively, removing reviews generally improves performance for zero-shot and DeepCoNN and generally insignificantly impacts performance for DIRECT, RGCL and REVLoRA, and generally slightly improves performance for few-shot. 
\end{mdframed}

\begin{figure*}[!h]
    \centering
     \subfloat{%
        \includegraphics[width=1\textwidth, trim=0 0 0 0, clip]{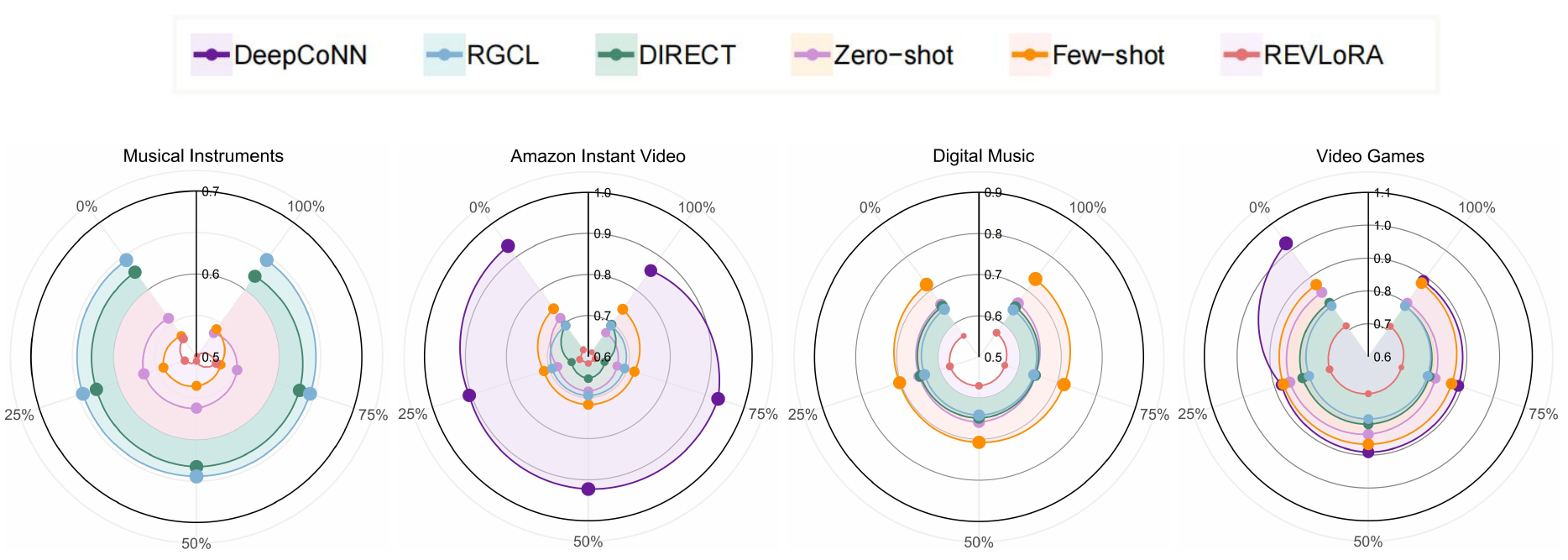}  
    }
    \vspace{-0.08in}
    \caption{
MAE comparison under varying review data reduction levels across four datasets. Each radar plot visualizes model performance at 0\%, 25\%, 50\%, 75\%, and 100\% review removal, with points closer to the center representing lower MAE. DeepCoNN is occasionally excluded for clarity due to its significantly lower performance. 
}
    \label{fig:reduction}
        \vspace{-0.15in}
\end{figure*}
\begin{figure*}[!h]
    \centering
     \subfloat{%
        \includegraphics[width=1\textwidth, trim=0 0 0 0, clip]{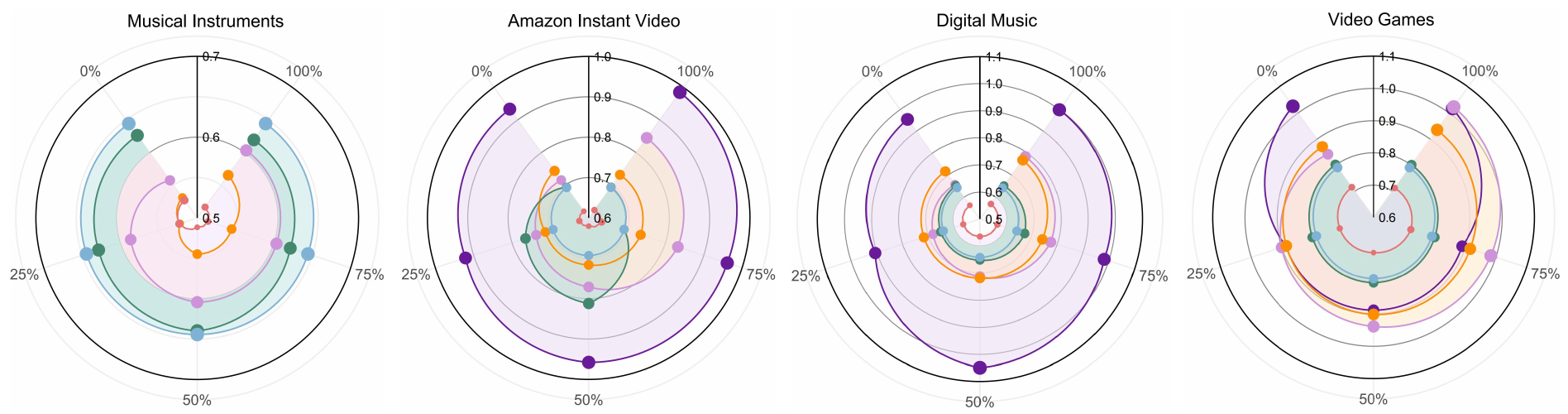}  
    }
        \vspace{-0.08in}
    \caption{
MAE comparison under varying review distortion levels across four datasets. Each radar plot visualizes model performance at 0\%, 25\%, 50\%, 75\%, and 100\% distortion, arranged in a counterclockwise direction. Points closer to the center indicate lower MAE. DeepCoNN is occasionally excluded for clarity due to its substantially degraded performance. 
}
    \label{fig:distortion}
    \vspace{-0.15in}
\end{figure*}
\begin{figure*}[!h]
    \centering
    \subfloat{%
        \includegraphics[width=1\textwidth, trim=0 0 0 0, clip]{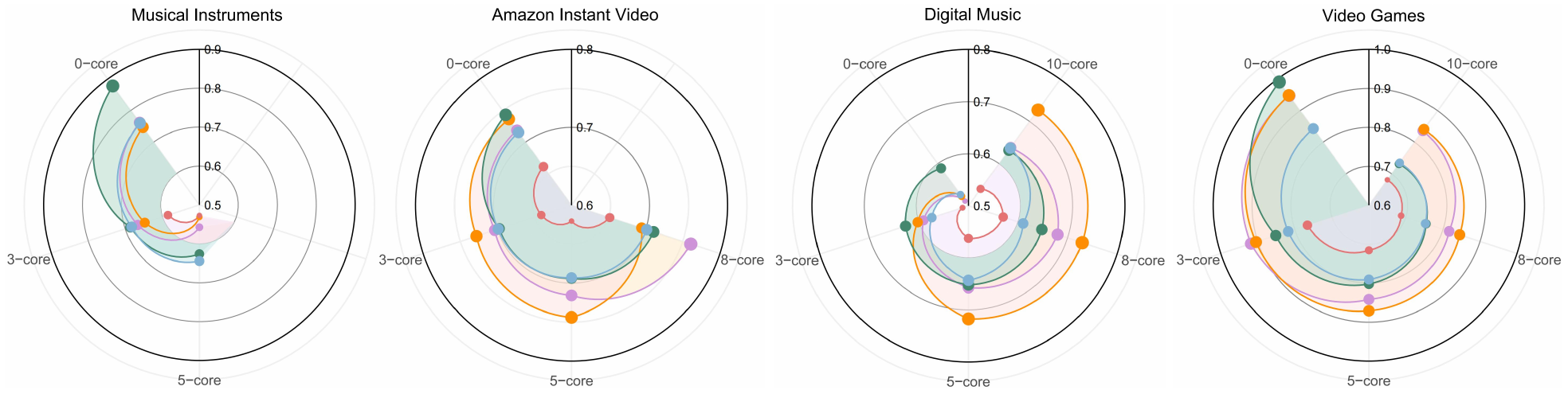}  
    }
        \vspace{-0.08in}
    \caption{MAE comparison under varying user–item interaction sparsity across four datasets. Each radar plot visualizes model performance at 0-, 3-, 5-, 8-, and 10-core levels (arranged counterclockwise), where higher $k$ indicates denser interaction filtering. Points closer to the center represent lower MAE. DeepCoNN is occasionally excluded for clarity due to its substantially degraded performance. 
}
    \label{fig:data-sparsity}
    \vspace{-0.15in}
\end{figure*}

\subsection{Reviews Reduction Experiment (RQ2)}
The reviews reduction experiments aim to examine model behavior under varying degrees of review reduction by randomly dropping a subset of reviews from the training set, while keeping rating data intact and leaving the validation and test sets unchanged. 
As shown in Fig. \ref{fig:reduction}, the model performance with partial removal is generally between the performance with 0\% review removal and the performance with 100\% review removal, up to variations due to random noise. For instance, the performance REVLoRA is generally consistent regardless of the degree of review reduction, while zero-shot LLM generally improves with increasing extent of reviews removed. DeepCoNN, occasionally not shown in Fig. \ref{fig:reduction} due to poor performance, shows improved performance as more reviews are removed, a continuation of the trend observed in the last subsection.

\begin{mdframed}[linecolor=black, backgroundcolor=gray!10, roundcorner=5pt, innertopmargin=5pt, innerbottommargin=5pt] \textbf{Answer to RQ2.}
A partial reduction in the number of available reviews generally results in a performance in between the scenarios where the reviews are completely removed, and where no review is removed.

\end{mdframed}

\subsection{Reviews Distortion Experiment (RQ3)}

While prior experiments examined the effects of removing review content, this subsection explores how textual distortion, rather than absence, impacts model performance. 
To simulate incoherent or mismatched contexts, we randomly select some or all review texts to shuffle among user–item pairs while preserving original rating interactions. This approach reflects real-world scenarios involving noisy, misaligned, or incorrectly attributed reviews. 
The results in Fig. \ref{fig:distortion} reveal significant differences in model robustness under review text distortion. 

The performance of DeepCoNN is relatively unstable, with MAE mostly increasing with increasing distortion. For example, in the Digital Music dataset, its MAE fluctuates from 0.9547 at 0\% distortion to 1.0495 at 50\%, then slightly recovers to 0.9990 at 100\%, indicating degraded performance under disrupted text–item alignment. DIRECT shows some performance degradation with smaller datasets under partial distortion, the most noticeable being Amazon Instant Video with 50\% distortion.
Zero-shot models degrade steadily with increasing distortion. For instance, in Video Games, MAE rises from 0.8417 to 1.0235, highlighting reliance on coherent review–item pairs. Few-shot models exhibit similar degradation but to a lesser extent, indicating partial resilience. 
Meanwhile, RGCL remains virtually unaffected across all datasets. Likewise, REVLoRA maintains stable performance across all examined distortion levels, with minor MAE variations. This suggests that these models either do not rely heavily on textual information or are robust to noise in review text. 

\begin{mdframed}[linecolor=black, backgroundcolor=gray!10, roundcorner=5pt, innertopmargin=5pt, innerbottommargin=5pt] \textbf{Answer to RQ3.}
Models differ in their sensitivity to review text distortion. While some models are resistant to distortion, other models may perform worse under partial or full distortion. Performance degradation may occur only in specific datasets, as is the case for DIRECT.

\end{mdframed}

\subsection{Data Sparsity Experiment (RQ4)}
Besides the availability and quality of review texts, data sparsity is a core challenge in recommender systems. To assess model robustness under varying sparsity levels, we conduct a $k$-core filtering experiment, retaining only users and items with at least $k$ interactions, and present their results in Fig. \ref{fig:data-sparsity}. 

DeepCoNN performs consistently poorly, with generally worse performance as $k$ increases, especially when $k \geq 3$. Due to its significantly degraded performance, we omit DeepCoNN from Fig. \ref{fig:data-sparsity} for visual clarity. Apart from DeepCoNN, DIRECT is more sensitive to data sparsity compared to the other frameworks examined. At 0-core, RGCL, zero-shot LLM and few-shot LLM consistently outperform DIRECT. The increase in MAE for DIRECT from 3-core to 0-core is consistently higher than other models across the examined datasets. However, as data becomes denser, DIRECT catches up and overtakes RGCL in most examined datasets. 

REVLoRA with MAE loss consistently outperforms across $k$-core datasets for $k \geq 3$ and 0-core Amazon Instant Video, highlighting its relative suitability for sparse data. 
Note that the 0-core experiment on REVLoRA is not completed for other datasets due to computational cost constraints, suggesting that inference costs for review summarization remain a significant computational bottleneck for large-scale datasets.

Generally, except DeepCoNN, as $k$ increases, the MAE of the model decreases. However, there are notable anomalies in Digital Music and the 8-core dataset of Amazon Instant Video. In these datasets, the most common rating is 5 but the proportion of reviews with rating 5 is generally lower at higher $k$, making the datasets more unpredictable. The increased unpredictability outweighs the increased data available to make rating predictions, leading to the anomaly.

\begin{mdframed}[linecolor=black, backgroundcolor=gray!10, roundcorner=5pt, innertopmargin=5pt, innerbottommargin=5pt] \textbf{Answer to RQ4.}
While performance generally improves with decreased data sparsity, lower sparsity does not guarantee lower MAE due to the variation of the rating distribution of the dataset when varying data sparsity. DIRECT and DeepCoNN are more sensitive to data sparsity than RGCL and LLM-based models. 
\end{mdframed}

\begin{figure*}[!ht]
    \centering
        \includegraphics[width=1\textwidth, trim=0 0 0 0, clip]{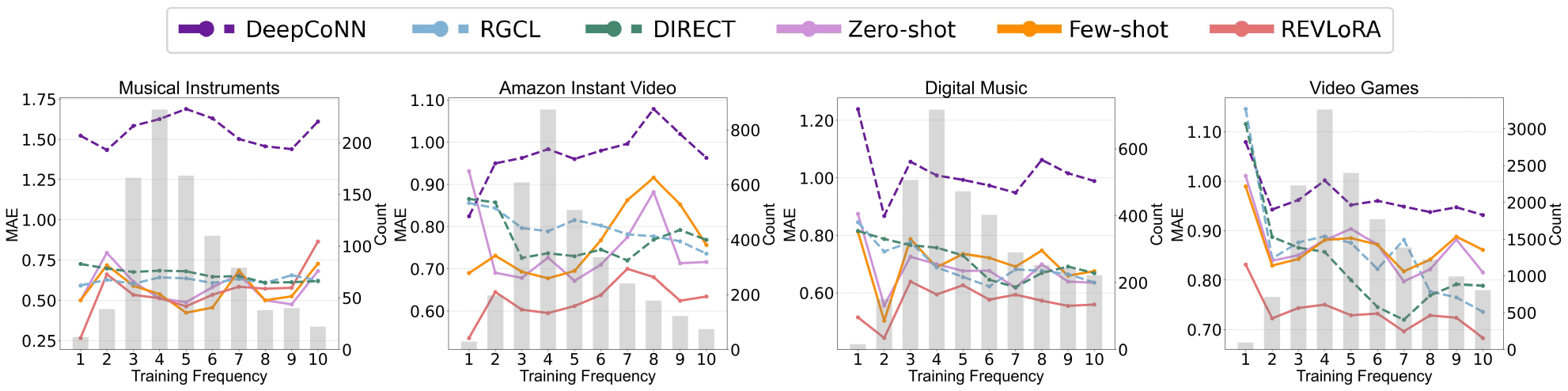} 

    \vspace{-0.15in}
        \caption{Cold-start MAE comparison across four Amazon datasets under different user training frequencies (1–10). 
Each line represents the MAE performance of a model when predicting ratings for test users grouped by the number of their past interactions in the training set. 
The gray bars (Count) indicate the number of test users in each training frequency group, i.e., how many users with $k$ interactions in the training set were evaluated at test time. 
This setup allows for a fine-grained analysis of model robustness under various cold-start conditions.
}
\vspace{-0.15in}
    \label{fig:cold-start}
\end{figure*}
\subsection{Cold-Start Experiment (RQ5)}

To assess model performance under cold-start conditions, we evaluate performance for users with limited historical interactions. 
Specifically, users are grouped by interaction frequency $f \in \{1, 2, \dots, 10\}$. For example, $f=1$ denotes users with a single training historical interaction. For each group, we independently compute MAE values and visualize the results in Fig. \ref{fig:cold-start}. 

DeepCoNN performs consistently poorly regardless of the datasets and the interaction frequency. In contrast, LLM-based models are generally more effective in sparse user scenarios. Zero-shot and few-shot LLM generally outperform deep learning baselines when $f \leq 5$, while the opposite is generally the case for $f > 5$. REVLoRA generally outperforms deep learning baselines across the interaction frequencies examines, with the outperformance margin generally larger when $f \leq 5$. This suggests that LLMs can generalize well from minimal user history by drawing on pre-trained linguistic, behavioral, and preference patterns learned from large-scale text corpora. Meanwhile, we observe that the performance of LLM solutions is highly correlated compared to deep learning solutions.

\begin{mdframed}[linecolor=black, backgroundcolor=gray!10, roundcorner=5pt, innertopmargin=5pt, innerbottommargin=5pt]
\textbf{Answer to RQ5.}  
In cold-start scenarios, LLM-based solutions generally outperform deep learning baselines. 
At interaction frequencies $f > 5$, deep learning methods generally outperform zero-shot and few-shot LLMs.
\end{mdframed}

\section{Conclusion}

In this paper, we evaluate the feasibility of using LLMs as review-aware rating prediction recommender systems and propose an evaluation framework, RAREval, that evaluates the sensitivity of review-aware recommender systems to varying degrees of removal and distortion of review text data, data sparsity in the dataset, and in cold-start scenarios.
We empirically find LLMs capable of delivering effective recommendations with textual reviews, and could achieve marginal improvement by further incorporating rating information (i.e., Few-shot Llama and REVLoRA). 
For both recent traditional and LLM-based rating prediction recommender systems, the addition of review text to review ratings demonstrated limited utility.
Future work could encompass prompt optimization for zero-shot, few-shot and finetuning, and the effectiveness of LLM in sequential review-aware rating prediction.

\bibliographystyle{IEEEtran}
\bibliography{ref}

\end{document}